\pdfoutput=1

\documentclass[11pt]{article}

\usepackage{ACL2023}

\usepackage{times}
\usepackage{latexsym}

\usepackage[T1]{fontenc}
\usepackage{amsfonts} 

\usepackage[utf8]{inputenc}

\usepackage{microtype}

\usepackage{inconsolata}
\usepackage{booktabs} 
\usepackage{multirow} 
\usepackage{graphicx}
\usepackage{amsbsy}
\usepackage{amsmath}
\usepackage{tabularx}

\title{Mitigating Hallucinations in LM-Based TTS Models via Distribution Alignment Using GFlowNets}

\author{Chenlin Liu\textsuperscript{1}, Minghui Fang\textsuperscript{2}, Patrick Zhang\textsuperscript{4}\thanks{\ \ Project Leader}, Wei Zhou\textsuperscript{2}, Jie Gao\textsuperscript{3}, Jiqing Han\textsuperscript{1}\thanks{\ \ Corresponding Author} \\
  \textsuperscript{1}Harbin Institute of Technology, China \\
  \textsuperscript{2}Zhejiang University, China \\
  \textsuperscript{3}Tsinghua University, Shenzhen, China \\
  \textsuperscript{4}Independent Researcher \\
  \texttt{chenlin.liu@stu.hit.edu.cn \quad jqhan@hit.edu.cn} \\}

\begin{document}
\maketitle
\begin{abstract}
Language Model (LM)-based Text-to-Speech (TTS) systems often generate hallucinated speech that deviates from input text. Existing mitigation strategies either demand excessive training resources or introduce significant inference latency. In this paper, we propose \textbf{G}Fl\textbf{O}wNet-guided distribution \textbf{A}lignmen\textbf{T} (GOAT) for LM-based TTS, a post-training framework that mitigates hallucinations without relying on massive resources or inference cost. Specifically, we first conduct an uncertainty analysis, revealing a strong positive correlation between hallucination and model uncertainty. Based on this, we reformulate TTS generation as a trajectory flow optimization problem and introduce an enhanced Subtrajectory Balance objective together with a sharpened internal reward as target distribution. We further integrate reward temperature decay and learning rate optimization for stability and performance balance. Extensive experiments show that GOAT reduce over 50\% character error rates on challenging test cases and lowering uncertainty by up to 58\%, demonstrating its strong generalization ability and effectiveness. Code: \href{https://github.com/lotuscarvedlife/GOAT}{https://github.com/lotuscarvedlife/GOAT}

\end{abstract}

\section{Introduction}
Text-to-Speech (TTS) aims to convert written text into high-fidelity speech and serves as a critical component in human-computer interaction \cite{kaur2023conventional, kumar2023deep}. Recently, advancements in large language models \cite{achiam2023gpt,guo2025deepseek,yang2025qwen3,grattafiori2024llama} and speech discretization techniques \cite{zhang2023speechtokenizer,lee2024high,huang-etal-2024-repcodec} have spurred the development of LM-based TTS models following the next-token prediction paradigm. These models \cite{wang2023neural,kharitonov2023speak,han2024vall,du2024cosyvoice,du2024cosyvoice2} sample the next token from the multinomial distribution conditional on previously generated tokens, which are very effective in modeling long sequences and can synthesize impressively high-quality speech.

\begin{figure}[t!]
    \centering
    \includegraphics[width=0.5\textwidth]{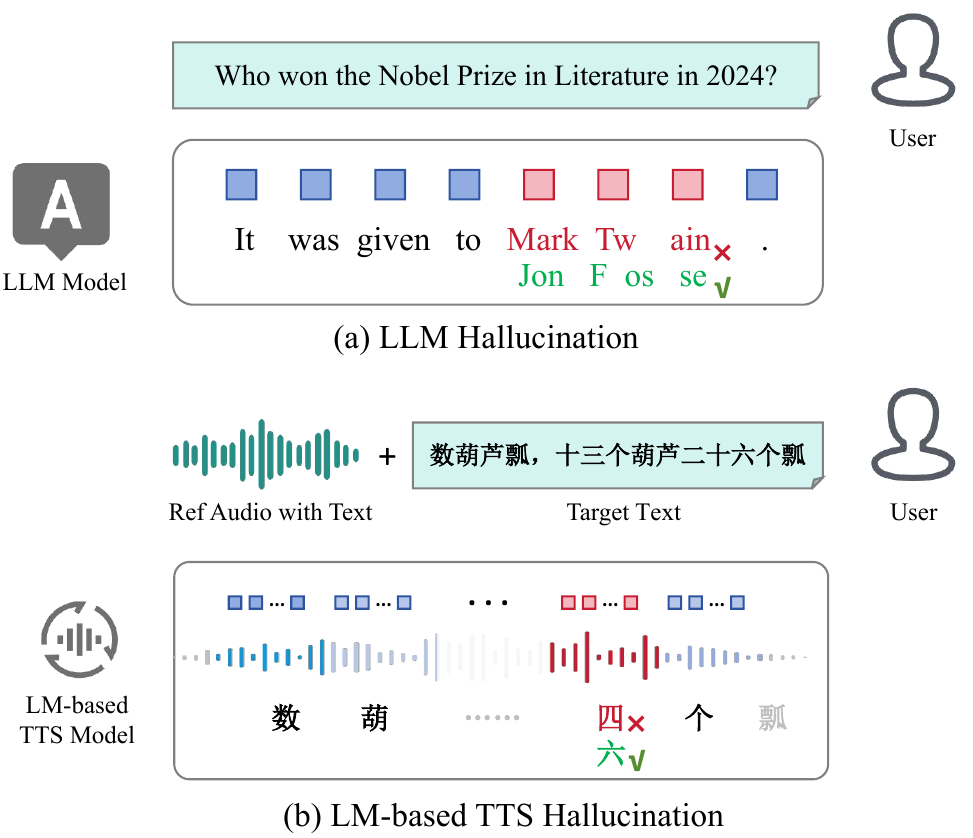}
    \caption{Hallucination phenomenon in LLM and LM-based TTS model. (a) LLM hallucination represents some wrong factual tokens. (b) LM-based TTS hallucination manifest as wrong pieces of token sequences, which often map to localized segments of generated audio containing inaccuracies such as mispronunciations, missing words, or semantic inconsistencies.}
    \label{fig:fig_hlc}
\end{figure}

Nevertheless, several studies \cite{ji2023survey, chuang2023dola} in the field of natural language processing have shown that language models are prone to hallucination, particularly when predicting tokens that convey factual information, as shown in Figure~\ref{fig:fig_hlc}. This phenomenon has unfortunately been inherited by LM-based TTS models, manifesting as generated speech that may deviate from the ground-truth text. Such issues become more pronounced when generating long or complex sentences, with errors such as mispronunciations, word omissions, and unnatural repetitions occurring more frequently.

To address these challenges, existing research has primarily focused on two directions: (1) \textbf{Training-time scaling:} Such approaches typically involve scaling up model parameters and increasing the size of corpus to develop more powerful TTS models \cite{peng2024voicecraft, ju2024naturalspeech, du2024cosyvoice2}. However, this incurs substantial computational costs and data collection burdens, particularly in resource-constrained scenarios. (2) \textbf{Test-time scaling:} Such approaches typically involve increasing test-time computation to enhance performance \cite{tu2024enabling, ye2025llasa}. However, the inference overhead presents significant challenges for real-time TTS applications, especially the slowed down generation speed. 

In light of this, we propose a \textbf{G}Fl\textbf{O}wNet-guided distribution \textbf{A}lignmen\textbf{T} framework (GOAT) for the post-training of LM-based TTS models. Specifically, we first comprehensively investigate the decoding process of LM-based TTS models, and identify potential correlations between TTS hallucinations and model uncertainty. Building on this observation, GOAT encourages the model to learn a probability distribution over state transitions, aiming to discover more deterministic and optimal decoding paths. GOAT tailors an internal reward distribution specifically for LM-based TTS models and achieves objective alignment through enhanced sub-trajectory balancing. To mitigate reward hacking, we carefully trade off model performance and training stability, and optimize the training process to ensure robust performance. GOAT enables intrinsic reinforcement learning without relying on large-scale training corpora or extensive computational resources, and does not introduce significant inference latency.

GOAT utilizes CosyVoice 2 \cite{du2024cosyvoice2} as its backbone and has undergone extensive evaluation in cross-lingual settings, demonstrating strong robustness and generalization capabilities. Further analysis from the perspective of information entropy validates the effectiveness of GOAT. GOAT achieves probability distribution alignment through intrinsic reinforcement learning, offering a novel perspective for the optimization of LM-based TTS models. Our contributions are highlighted as follows.

\begin{itemize}
    \item To the best of our knowledge, GOAT is the first work that leverages GFlowNet to optimize speech synthesis through distribution alignment.
    
    \item GOAT provides the first in-depth investigation of the weaknesses of LM-based TTS models from the model uncertainty perspective, which provides insights for subsequent work.

    \item GOAT tailors the reward function as well as the optimization strategy for the LM-based TTS model, and comprehensive experiments demonstrate its effectiveness.
    
\end{itemize}

\section{Hallucination Analysis in LM-based TTS models}
\label{sec:h_analysis}
We begin with a comprehensive analysis of hallucination issues in LM-based TTS models (such as mispronunciations, omissions, and unnatural repetitions), which serves as the key motivation and theoretical foundation for GOAT.

\subsection{Hallucination Detection}
\label{sec:hallucination_detection_methods}
Recent studies \cite{huang2024survey, ma2025estimating} have analyzed the hallucination problem of language models in text-based tasks from the perspective of model uncertainty. Inspired by this, we adopt the entropy-based metric for speech hallucination detection. More detailed discussion can be found in Appendix~\ref{sec:Modality-Specific Hallucination Detection in Speech}. Given that speech tokens have significantly lower information density compared to text tokens, multiple tokens must be aggregated to represent the pronunciation of a single character. With this in mind, we assess the uncertainty of LM-based TTS models across three dimensions: token level, character level, and utterance level.

Specifically, for the token probability distribution $P_{t} = \{p_1, \ldots, p_{|C|}\}$ by the model at time step $t$, the token-level uncertainty can be defined as:
\begin{align}
   \label{eq:entropy}
    H_{\text{token}}(P_{t}) = -\sum_{i=1}^{|C|} p_i \log p_i 
\end{align}
where $|C|$ denotes the vocabulary size. We extract the left and right boundaries $i$ and $j$ of each character $W$ within the token sequence to compute the character-level uncertainty $H_{\text{character}}(W_{i:j})$, while the utterance-level uncertainty $H_{\text{utterance}}(S)$ is defined as the average uncertainty over all tokens. They can be formally defined as follows:
\begin{align}
    H_{\text{character}}(W_{i:j}) &= \frac{1}{j-i} \sum_{k=i}^j H_{\text{token}}(P_{k}) \label{eq:word_uncertainty} \\
    H_{\text{utterance}}(S) &= \frac{1}{|S|} \sum_{k=1}^{|S|} H_{\text{token}}(P_{k}) \label{eq:utterance_uncertainty}
\end{align}
where $|S|$ denotes the sequence length. More discussion can be found in Appendix~\ref{appendix:hdm}. Intuitively quantifying model uncertainty facilitates the exploration of its potential connection to hallucination.

\subsection{Empirical Analysis}
\label{sec:utterance-level_analysis}
We conducted experiments on SeedTTS-Eval benchmark \cite{anastassiou2024seed} utilizing CosyVoice 2 \cite{du2024cosyvoice2}. To reveal potential hallucinations, we select 200 samples from the \textit{test-hard} subset and generate speech using stochastic multinomial sampling. Paraformer-zh \cite{gao2022paraformer} is employed to perform automatic speech recognition (ASR) to compute the character error rate (CER), which serves as the hallucination proxy. 

\begin{figure}[t!]
    \centering
    \includegraphics[width=1\linewidth]{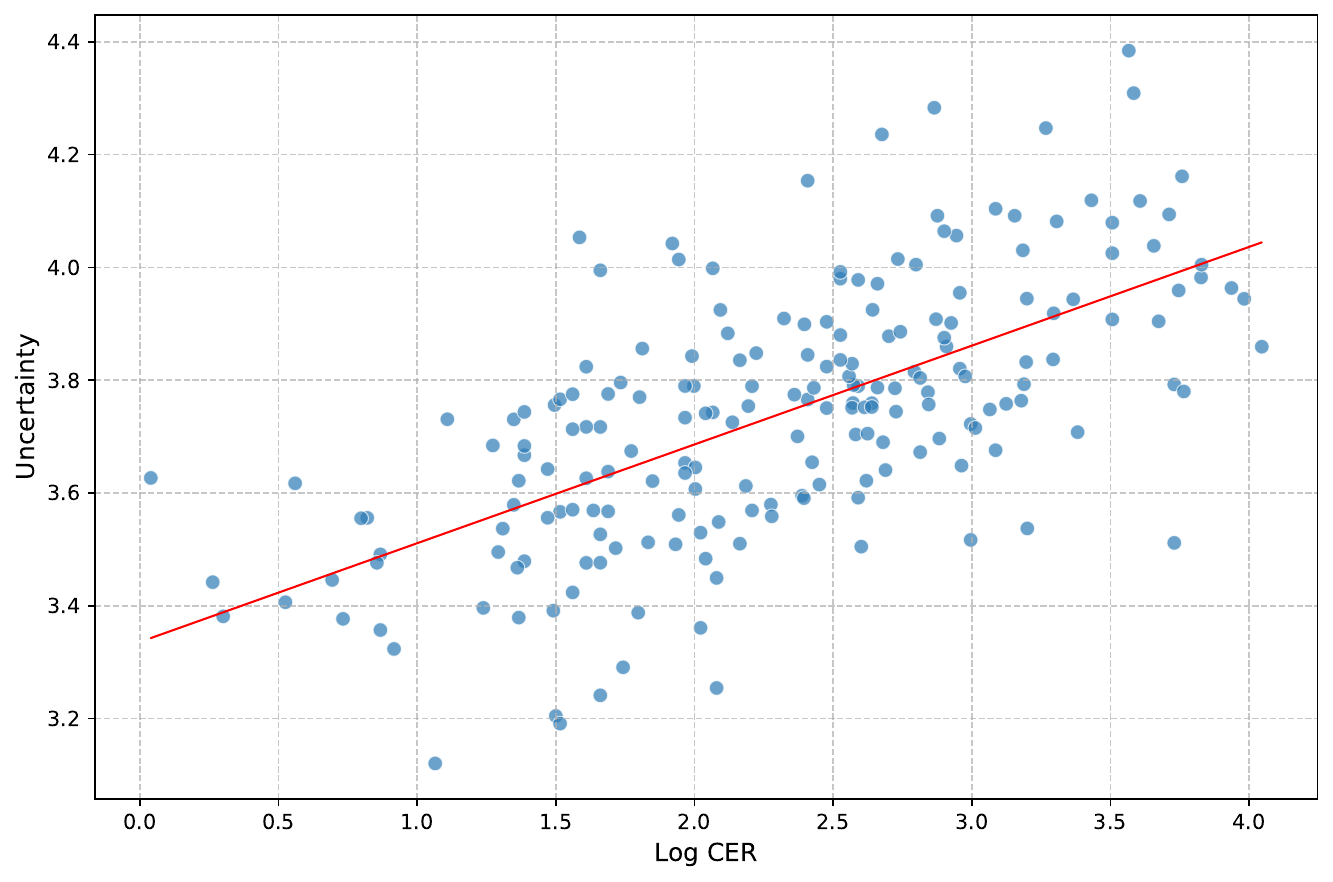}
    \caption{The relationship between utterance-level uncertainty and log CER. The red line represents the linear regression fit.}
    \label{fig:uncertainty_wer_correlation}
\end{figure}

As shown in Figure~\ref{fig:uncertainty_wer_correlation}, we annotate the utterance-level uncertainty and the corresponding CER for all samples, and perform linear regression. We observe that utterances with low CER typically exhibit lower uncertainty, and vice versa. To enhance the credibility of the results, we further compute the Pearson correlation coefficient and the Spearman rank correlation coefficient, which are 0.636 and 0.649 respectively ($p<1\mathrm{E}-4$), revealing statistically significant positive correlations. More discussion and detailed analysis of token-level/character-level uncertainty is provided in Appendix~\ref{appendix:fine-grained_analysis}.

Given the positive correlation between model uncertainty and hallucination generation, GOAT encourages the model to discover more deterministic and optimal decoding paths to enhance the performance of LM-based TTS models.

\section{GOAT}
GOAT tailors an internal reward distribution specifically for LM-based TTS models and achieves objective alignment through enhanced sub-trajectory balancing. Building on the aforementioned findings, GOAT leverages GFlowNet to adjust the probability distribution at the sequence level, aiming to explore more deterministic and optimal decoding paths. The intuition behind this is that GOAT assumes the LM can reliably assign a high likelihood to high-probability sentences, and wishes to preferentially sample over low-probability ones.

\subsection{Adapting GOAT to LM-based TTS}
We present the core pipeline and formulas of GFlowNets. The detailed theoretical foundations are provided in Appendix~\ref{sec:appendix_gflownets}, which we strongly recommend readers consult to deepen understanding. 
GFlowNets encourage the model to learn an optimal transition process from the initial state $s_0$ to the terminal state $s_n$, where the path from $s_0$ to $s_n$ forms a trajectory $\tau$. The forward policy $P_F(\tau)$ of GFlowNets determines a distribution over $\tau$ by:
\begin{align}
    P_F(\tau = (s_0 \to \cdots \to s_n)) = \prod_{i=0}^{n-1} P_F(s_{i+1} | s_i).
\end{align}

Each state transition is associated with a reward. Given reward function $R(x)$ defined over the set of terminal states $\mathcal{X}$, the forward policy $P_F(\tau)$ of GFlowNets should be proportional to it: 
\begin{align}
\label{eq:reward_proportionality2}
    R(x) = Z \sum_{\substack{\tau = (s_0 \to \cdots \to s_n = x)}} P_F(\tau) \quad \forall x \in \mathcal{X},
\end{align}  
where the normalization constant $Z = \sum_{x \in \mathcal{X}} R(x)$. To align with the reward distribution, GFlowNet brings in the trajectory flow  $F(\tau) \propto P_F(\tau)$, which is an unnormalized probability function. By summing the trajectory flows $F(\tau)$ over all trajectories $\tau$ that terminate at state $x$, the state flow $F(x)$ can be obtained:

\begin{align}
F(x) = \sum_{x \in \tau} F(\tau).
\end{align}  

GFlowNets follow the principle of flow conservation, which for any terminal state $x$: 
\begin{align}
\label{eq:flow_normalization}
    F(x) = R(x) \quad \forall x \in \mathcal{X}.
\end{align} 

GOAT treats the autoregressive generation process of LM-based TTS models as a sequence-level state transition process. GOAT applies the forward sampling policy $P_{\text{GFN}}$ to generate complete token sequences $\mathbf{a}^\top = a_1 \dots a_n\top \in \mathcal{X}$, where $\top$ denotes the terminal token. The process is as follows:

\begin{enumerate}
    \item \textbf{Initial Setup}: Given a conditioning sequence $\mathbf{q}$ and model parameters $\theta$, the initial state $s_0$ is an empty token sequence denoted as $\mathbf{a}_0$.
    \item \textbf{State Transition} At decoding step $t$, the next token is sampled from $P_{\text{GFN}}(s_t \mid s_{t-1}, \mathbf{q}, \theta)$, where $s_{t-1} = \mathbf{a}_{t-1} = a_1 \dots a_{t-1}$. The next state is then updated to $s_t = \mathbf{a}_t = a_1 \dots a_t$.
    \item \textbf{Termination}: The pipeline stops upon sampling the termination token $\top$, yielding a complete token sequence $\mathbf{a}^\top \in \mathcal{X}$.
\end{enumerate}

\begin{figure*}[t!]
    \centering
    \includegraphics[width=0.9\textwidth]{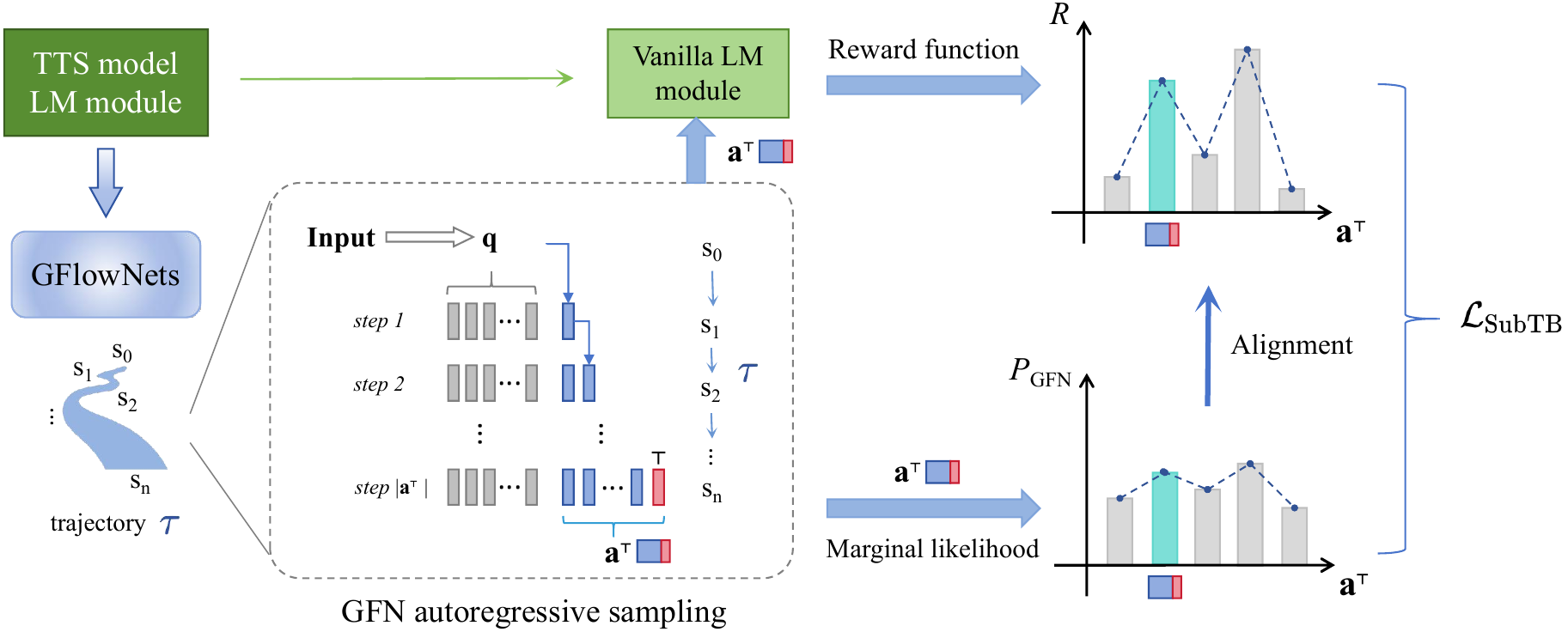}
    \caption{Training pipeline of GOAT.}
    \label{fig:pipeline}
\end{figure*}

This entire process defines a complete trajectory $\tau = \mathbf{a}_0 \to \dots \to \mathbf{a}^\top$, as shown in Figure~\ref{fig:pipeline}, with the trajectory probability distribution given by:  
\begin{equation}
\begin{aligned}
\label{eq:GFN_to_token_sequence}
    P_{\text{GFN}}(\tau) 
    &= \prod_{i=1}^{|\mathbf{a}^\top|} P_{\text{GFN}}(s_i \mid s_{i-1}, \mathbf{q}, \theta) \\
    &= \prod_{i=1}^{|\mathbf{a}^\top|} P_{\text{GFN}}(a_i \mid a_{1:i-1}, \mathbf{q}, \theta), \\
\end{aligned}
\end{equation}
where $|\mathbf{a}^\top| = n$ is the length of the complete sequence, and $a_{1:0} = \mathbf{a}_0$. LM-based TTS models follow the next-token prediction paradigm, where only one unique path leads to the next state given the fixed prefix. Therefore, for any terminal state $x = \mathbf{a}^\top$, there exists only one trajectory $\tau$. Thus, the marginal likelihood of $\mathbf{a}^\top$ is: 

\begin{equation}
\begin{aligned}
    P_{\text{GFN}}^\top(\mathbf{a}^\top) 
    &= \sum_{\mathbf{a}^\top \in \tau} P_{\text{GFN}}(\tau) \\
    &= \prod_{i=1}^{|\mathbf{a}^\top|} P_{\text{GFN}}(a_i \mid a_{1:i-1}, \mathbf{q}, \theta).
\end{aligned}
\end{equation}
Based on this, the objective of GOAT is to learn a forward sampling policy $P_{\text{GFN}}(\cdot \mid \cdot, \mathbf{q}; \theta)$ that aligns with distribution of reward function  $R: \mathcal{X} \to \mathbb{R}_{\ge 0}$. Formally:

\begin{align}
    P_{\text{GFN}}^\top(\mathbf{a}^\top) \propto R(\mathbf{a}^\top) \quad \forall \mathbf{a}^\top \in \mathcal{X}.
\end{align}

\subsection{Training Objective}

\subsubsection{Enhanced Subtrajectory Balance}
Our empirical analysis reveals that LM-based TTS models tend to exhibit fragmentary collapse when generating long and complex sentences. This property necessitates that GOAT perform fine-grained optimization. To this end, we employ the SubTrajectory Balance (SubTB) loss \cite{madan2023learning} for training, which enables the model to learn from subsequences of varying lengths. For any subtrajectory $\tau_{m:n} = (s_m \to \cdots \to s_n)$, the SubTB loss can be defined as:
\begin{equation}
\begin{aligned}
    \label{eq:subtb_loss}
    &\mathcal{L}_{\text{SubTB}}(\tau_{m:n}, \mathbf{q}; \theta) = \\
    &\left( \log \frac{F(s_m) \prod_{i=m}^{n-1} P_F(s_{i+1} \mid s_i, \mathbf{q}, \theta)}{F(s_n) \prod_{i=m}^{n-1} P_B(s_i \mid s_{i+1}, \mathbf{q}, \theta)} \right)^2,
\end{aligned}
\end{equation}
where $P_F$ denotes the forward policy for generating the next state, while $P_B$ denotes the backward policy for tracing back to the previous state based on the posterior distribution.

In LM-based TTS models, the autoregressive generation process imposes a deterministic prefix on the current sequence, such that each state has one unique parent state, which implies $P_B \equiv 1$. Following Equation~\ref{eq:flow_normalization}, we replace the state flow $F$ in Equation~\ref{eq:subtb_loss} with the reward function $R(\mathbf{a}^\top)$. This allows the model to optimize the forward sampling policy $P_F$ directly, without the need to parameterize $F$. Finally, by incorporating all subtrajectories under the constraint of Equation~\ref{eq:GFN_to_token_sequence}, the final training objective is defined as follows:
\begin{equation}
\begin{aligned}
    &\mathcal{L}(\mathbf{a}^\top, \mathbf{q}; \theta) = 
    \sum_{0 \leq i \leq j \leq |\mathbf{a}^\top|} \\ &\left( \log \frac{R(\mathbf{a}_i) \prod_{k=i}^{j-1} P_{\text{GFN}}(a_{k+1} \mid a_{1:k}, \mathbf{q}, \theta)}{R(\mathbf{a}_j)} \right)^2,
\end{aligned}
\end{equation}
where $\mathbf{a}_i, \mathbf{a}_j$ represent partial sequences at positions $i, j$.  

\subsubsection{Internal Reward}
\label{sec:reward_function}
To reduce resource dependency, we have meticulously designed an intrinsic reward function for GOAT. This enables GOAT to be trained on unlabeled data without relying on external reward models. Specifically, in the autoregressive generation of LM-based TTS models, the normalized token sampling probability $p_{\text{LM}}(\cdot \mid \cdot)$ at each generation step serves as the most intuitive reward signal. By accumulating all tokens rewards in the subsequence $\mathbf{a}_k$, we define the reward for $\mathbf{a}_k$ as:
\begin{align}
    R(\mathbf{a}_k |\mathbf{q}) = p_{\text{LM}}(\mathbf{a}_k|\mathbf{q})
\end{align}
This reward function maintains implicit alignment with the original sequence distribution of the backbone model. Furthermore, learning a more concentrated reward distribution contributes to reducing model uncertainty. To achieve this, we apply an inverse temperature $ T $ (where $ 0 < T < 1 $) to the sampling probabilities in order to sharpen the reward distribution. The final reward function is defined as follows:
\begin{equation}
\begin{aligned}
    R(\mathbf{a}_k| \mathbf{q}) &= p_{\text{LM}}(\mathbf{a}_k \mid \mathbf{q})^{\frac{1}{T}} \\
    &= \left( \prod_{i=1}^k p_{\text{LM}}(a_i \mid a_{1:i-1}, \mathbf{q}) \right)^{\frac{1}{T}},
\end{aligned}
\end{equation}
This reward function is grounded in the GOAT hypothesis, which posits that language models inherently assign higher probabilities to speech token sequences of superior quality. GOAT enhances model determinism through sequence rewards, encouraging the model to prioritize sampling high-probability sequences over low-probability ones.

It is worth noting that the temperature strategy proposed by GOAT is distinct from the low-temperature sampling \cite{brown2020language} typically applied during token generation. The temperature strategy of GOAT influences the reward distribution across the entire sequence, encouraging the model to favor globally optimal sequences. In contrast, low-temperature sampling focuses on adjusting the probability distribution at each decoding step, achieving only locally optimal outcomes. As the temperature $T$ in low-temperature sampling approaches 0, the probability of high-probability tokens tends toward positive infinity, causing the sampling process to degenerate into greedy sampling. Consequently, this fails to ensure the generation of high-quality complete sequences.

\subsection{Reward Hacking Suppression}
\label{sec:reward_hacking_suppression}

Reward hacking \cite{amodei2016concrete,pan2022effects} refers to the phenomenon where the model takes actions that are misaligned with the intended task to steal rewards. In GOAT, reward hacking is observed when the model generates speech token sequences with abruptly shortened lengths, leading to premature termination of audio waveforms. To mitigate this, we have carefully refined the training strategy.


\textbf{Reward Temperature Decay. }
Ideally, a lower reward temperature yields a sharper reward distribution, which can enhance model performance. However, it also significantly increases the susceptibility of reward hacking. To balance performance and stability, we impose a linearly decaying reward temperature during training, starting from 1 and decreasing to a predefined lower bound. This reduces instability at the beginning of training, while gradually guiding the model toward more optimal solutions.

\textbf{Learning Rate Optimization.}
Given that an excessively high learning rate may cause the model to learn anomalous and short-sighted high reward behaviors, we design a stabilized learning rate optimization scheme. Specifically, we employ the combination of warm-up and cosine annealing to ensure a more effective optimization trajectory. These modifications significantly reduced reward hacking occurrences while maintaining model performance.

\section{Experiments}
\label{sec:experiments}
\subsection{Datasets \& Baseline}
\begingroup
\textbf{Datasets } We randomly select 1000 samples from LibriTTS \cite{zen2019libritts} and WenetSpeech4TTS \cite{ma2024wenetspeech4tts} for training on Chinese and English. For evaluation, we use the SeedTTS-Eval benchmark \cite{anastassiou2024seed}, with further details provided in Appendix~\ref{appendix:datasets}.

\noindent \textbf{Baseline } We use a standard LM-based TTS architecture, CosyVoice 2 \cite{du2024cosyvoice}, as the foundation model for GOAT.
\endgroup

\subsection{Implementation Details}
We initialize the GFlowNet with the original language model and subsequently post-train it using Low-rank Adaptation (LoRA) \cite{hu2022lora}. We use 4 NVIDIA H100 Hopper GPUs for model training and one NVIDIA V100 GPU for evaluation. The implementation includes two configurations balancing performance and training stability, with details provided in Appendix~\ref{appendix:implementation_details}.

\subsection{Sampling Methods}
\label{sec:sampling_methods}
We adopt the Repetition Aware Sampling (RAS) strategy from CosyVoice 2 under its default hyperparameters (see Appendix~\ref{appendix:implementation_details} for details). For a more fine-grained comparison, we introduce random multinomial sampling (RMS). Additionally, to verify the difference discussed in Section~\ref{sec:reward_function}, we also include a baseline that applies the same low temperature as used in the reward function.



\subsection{Experimental Results}
\label{sec:experimental_results}

\subsubsection{Metrics}
We assess the synthesized speech based on content consistency (Character Error Rate \& Word Error Rate, CER/WER), Speaker Similarity (SS), and speech quality (UTMOS, \cite{saeki2022utmos}). More details can be found in Appendix~\ref{appendix:metrics}.

\begin{table*}
\small
\centering
\setlength{\tabcolsep}{2pt} 
\begin{tabular*}{\linewidth}{@{\extracolsep{\fill}}l l c c c c c c c c c}
\toprule
\multirow{2}{*}{Model} & \multirow{2}{*}{SM} 
& \multicolumn{3}{c}{\textit{test-zh}} 
& \multicolumn{3}{c}{\textit{test-en}} 
& \multicolumn{3}{c}{\textit{test-hard}} \\
\cmidrule(lr){3-5} \cmidrule(lr){6-8} \cmidrule(lr){9-11}
& 
& CER (\%) $\downarrow$ & SS $\uparrow$ & UTMOS $\uparrow$ 
& WER (\%) $\downarrow$ & SS $\uparrow$ & UTMOS $\uparrow$ 
& CER (\%) $\downarrow$ & SS $\uparrow$ & UTMOS $\uparrow$ \\
\midrule
Human & -- & 1.31 & 0.78 & 2.785 & 2.74 & 0.82 & 3.536 & {-} & {-} & {-} \\
\midrule
baseline & RMS & 4.61 & 0.84 & 3.102 & 7.10 & 0.78 & 3.880 & 13.72 & 0.82 & 2.849 \\
& RAS & 1.36 & 0.84 & 3.280 & 3.35 & 0.79 & 4.099 & 8.16 & 0.83 & 3.115 \\
& LT-RMS & 2.69 & 0.84 & 3.219 & 4.63 & 0.78 & 4.023 & 9.82 & 0.83 & 3.021 \\
& LT-RAS & 1.31 & 0.84 & 3.308 & 3.31 & 0.78 & 4.124 & 8.25 & 0.82 & 3.168 \\
\midrule
CV2-GOAT-zh(1500S) & RMS & 0.94 & 0.83 & 3.387 & 2.43 & 0.80 & 4.170 & 6.61 & 0.81 & 3.254 \\
& RAS & 0.89 & 0.83 & 3.401 & 2.02 & 0.80 & 4.170 & 6.28 & 0.81 & 3.255 \\
\midrule
CV2-GOAT-zh(2500S) & RMS & 0.90 & 0.82 & 3.394 & 2.27 & 0.80 & 4.200 & 6.53 & 0.81 & 3.273 \\
& RAS & 0.85 & 0.82 & 3.401 & 1.96 & 0.80 & 4.216 & 6.36 & 0.80 & 3.267 \\
\midrule
CV2-GOAT-en(1500S) & RMS & 1.26 & 0.84 & 3.300 & 2.16 & 0.80 & 4.184 & 7.40 & 0.82 & 3.146 \\
& RAS & 1.04 & 0.84 & 3.336 & 2.07 & 0.81 & 4.196 & 6.56 & 0.82 & 3.177 \\
\midrule
CV2-GOAT-en(2500S) & RMS & 1.16 & 0.84 & 3.299 & 2.13 & 0.81 & 4.182 & 7.37 & 0.82 & 3.131 \\
& RAS & 0.96 & 0.84 & 3.329 & 2.16 & 0.81 & 4.204 & 6.76 & 0.82 & 3.160 \\
\midrule
CV2-GOAT-mix(1500S) & RMS & 0.98 & 0.84 & 3.385 & 2.18 & 0.80 & 4.209 & 6.54 & 0.82 & 3.256 \\
& RAS & 0.88 & 0.83 & 3.389 & 2.06 & 0.80 & 4.225 & 6.51 & 0.82 & 3.277 \\
\midrule
CV2-GOAT-mix(2500S) & RMS & 0.86 & 0.83 & 3.386 & 2.13 & 0.80 & 4.211 & 6.55 & 0.82 & 3.253 \\
& RAS & 0.88 & 0.83 & 3.388 & 2.08 & 0.80 & 4.227 & 6.56 & 0.81 & 3.272 \\
\bottomrule
\end{tabular*}
\caption{\label{tab:results_full} Evaluation results across models. CV2-GOAT-zh/en/mix: models fine-tuned on Chinese/English/Mix of two datasets; 1500S/2500S : total steps in Learning Rate Optimization. SM: Sampling Method; LT prefix: sampling method using Low-Temperature.}
\end{table*}

\subsubsection{Comparison with Baseline}
\label{sec:results_compared_with_baseline}

As shown in Table \ref{tab:results_full}, models trained and evaluated on the same language outperform or achieve comparable performance to the baseline and even ground truth across all metrics. Notable improvements are observed in WER/CER and UTMOS scores, particularly with a more than 50\% reduction in WER/CER under RMS on the hallucination-prone \textit{test-hard} subset. Regarding SS, perceptual differences are minimal when SS values are sufficiently high (e.g., above 0.7) \cite{tu2024enabling}. These results strongly demonstrate the efficacy of GOAT in mitigating hallucinations.



\subsubsection{Comparison with Low-temperature Sampling}
As shown in Table~\ref{tab:results_full}, while low-temperature sampling achieves partial hallucination suppression, GOAT significantly outperforms this strategy, surpassing over 30\% on CER/WER. This provides compelling evidence supporting the distinction between GOAT and low-temperature sampling presented in Section~\ref{sec:reward_function}. 

\subsubsection{Generalization and Effectiveness}

For models trained and evaluated on different languages, results in Table \ref{tab:results_full} still outperform the baseline, only marginally inferior to models trained on matched-language data. For models trained on mixed data, despite halving data volume for each language, the model achieved comparable performance on both languages. These observations demonstrate the strong robustness and generalization capability of GOAT.

Intriguingly, the marginal gains observed when applying RAS sampling suggest that GOAT effectively steers probability distributions toward higher-quality speech token sequences.

\subsubsection{Ablation Experiments}

\begin{table}[t!]
    \small
    \centering
    \begin{tabular}{l l c c c}
        \toprule
        Loss & LRO Steps & CER (\%) $\downarrow$ & SS $\uparrow$ & UTMOS $\uparrow$ \\
        \midrule
        baseline & -- & 13.72 & 0.82 & 2.849 \\
        \midrule
        TB & 1500 & 11.72 & 0.82 & 2.923 \\
        & 2500 & 11.84 & 0.82 & 2.94 \\
        \midrule
        SubTB & 1500 & 6.61 & 0.81 & 3.254 \\
        & 2500 & 6.53 & 0.81 & 3.273 \\
        \bottomrule
    \end{tabular}
    \caption{\label{tab:loss_comparison} Comparison of TB and SubTB using the same 
 enhancement method, evaluated on \textit{test-hard}}
\end{table}

Furthermore, we conducted ablation experiments on each strategy that could be removed. The results are as follows:

\begin{figure*}[t!]
    \centering
    \includegraphics[width=1\textwidth]{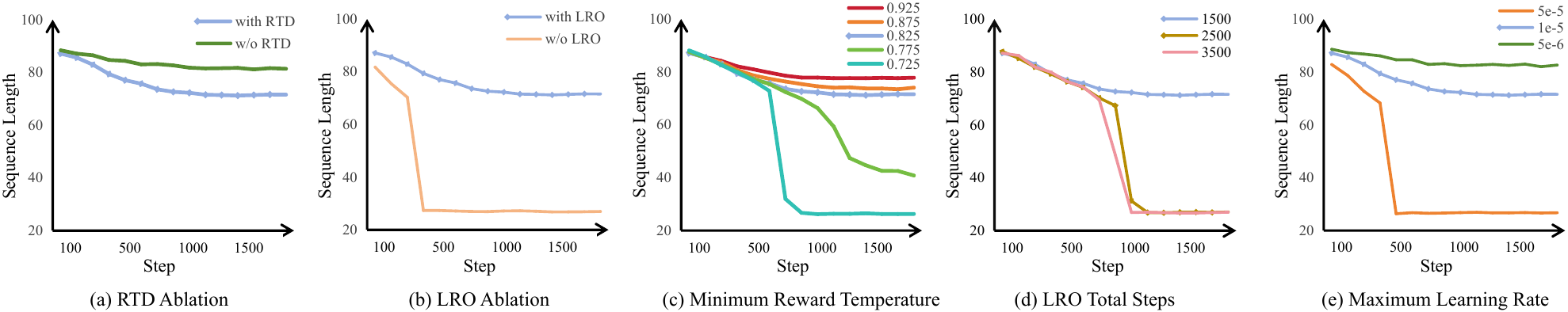}
    \caption{Evaluation of different configuration. Use line with nodes to represent our configuration.}
    \label{fig:reward_hacking_ablation}
\end{figure*}

\noindent \textbf{w/o enhanced SubTB } In this setting, SubTB is degraded to Trajectory Balance \cite{malkin2022trajectory} with the same enhancement. As shown in Table \ref{tab:loss_comparison}, we observe that without the ability to learn from subsequences of varying lengths, the model does not effectively mitigate hallucinations.

\noindent \textbf{w/o reward temperature decay } This setting removes reward temperature decay (RTD), directly using the minimum temperature for training. The performance is shown in Figure \ref{fig:reward_hacking_ablation} (a). Despite a seemingly more stable training process, it does not lead to better convergence, with performance details provided in Appendix~\ref{appendix:model_p_under_hyperp_config}.

\noindent \textbf{w/o learning rate optimization } In this configuration, learning rate optimization (LRO) is omitted, and the training uses a fixed learning rate. The performance is shown in Figure \ref{fig:reward_hacking_ablation} (b), where the model is more prone to reward hacking, leading to a sudden drop in the output sequence length. For performance details, please refer to Appendix~\ref{appendix:model_p_under_hyperp_config}

\subsection{In-depth Analysis}

\begin{figure}[t!]
    \centering
    \includegraphics[width=0.85\linewidth]{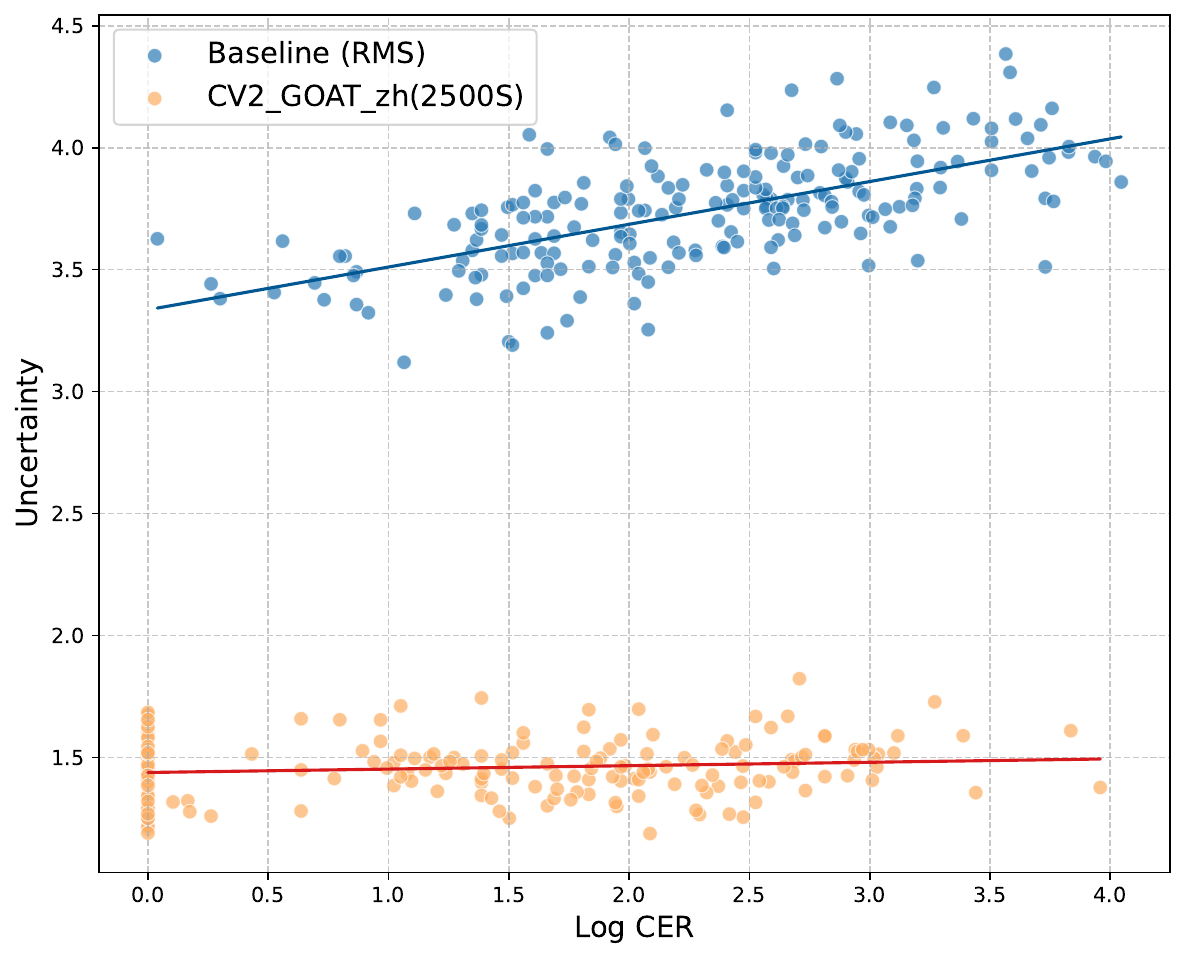}
    \caption{Comparison of the utterance-level uncertainty and log CER between the two models. For those who have zero CER, we set the log CER to 0.}
    \label{fig:comparison_of_uncertainty_analysis}
\end{figure}

\begin{table}[t!]
    \small
    \centering
    \begin{tabular}{l c c c}
        \toprule
        \multirow{2}{*}{Model} & \multicolumn{3}{c}{Average UUR} \\
        \cmidrule(lr){2-4}
        & \textit{test-zh} & \textit{test-en} & \textit{test-hard} \\
        \midrule
        baseline & 1.00 & 1.00 & 1.00 \\
        CV2-GOAT-zh(1500S) & 0.53 & 0.66 & 0.50 \\
        CV2-GOAT-zh(2500S) & 0.42 & 0.58 & 0.39 \\
        CV2-GOAT-en(1500S) & 0.67 & 0.56 & 0.63 \\
        CV2-GOAT-en(2500S) & 0.66 & 0.54 & 0.62 \\
        CV2-GOAT-mix(1500S) & 0.56 & 0.59 & 0.52 \\
        CV2-GOAT-mix(2500S) & 0.47 & 0.51 & 0.44 \\
        \bottomrule
    \end{tabular}
    \caption{\label{tab:uur_results}Comparison of average UUR across models. Set baseline model as benchmark.}
\end{table}

\textbf{Uncertainty Analysis } We also evaluate the effectiveness of GOAT from the perspective of uncertainty. 
Leveraging the utterance-level uncertainty defined in Formula \ref{eq:utterance_uncertainty}, we compare CV2-GOAT-zh(2500S) with the baseline as shown in Figure~\ref{fig:uncertainty_wer_correlation}. The results in Figure~\ref{fig:comparison_of_uncertainty_analysis} reveal that the aligned model exhibits significantly lower utterance-level uncertainty than the baseline. Additionally, the overall data cluster shifts leftward, indicating reduced CER (as confirmed in Section~\ref{sec:results_compared_with_baseline}), demonstrating effective hallucination suppression.

To quantify uncertainty reduction, we define the Utterance Uncertainty Ratio (UUR) as
\begin{equation}
    \text{UUR} = \frac{1}{N} \sum_{i=1}^N \frac{\sigma_{\text{trained},i}}{\sigma_{\text{baseline},i}},
\end{equation}
where $\sigma$ is the utterance-level uncertainty and N is the number of test utterances. Table~\ref{tab:uur_results} summarizes UURs across all models and evaluation sets. All models achieve substantial uncertainty reduction by up to 58\%. And both Chinese and English models exhibit consistent improvements even on the evaluation datasets with unseen language. Models trained on mixed data showing comparable performance on both language demonstrating the efficacy and generalization of GOAT.


\noindent \textbf{Hyperparameters Configuration } 
We investigate the impact of several key hyperparameters on model training stability and performance. As shown in the results in Figure~\ref{fig:reward_hacking_ablation}, all three hyperparameters play a crucial role in suppressing reward hacking, and our proposed settings balance training stability and model performance. An analysis of  model performance with various configuration is provided in Appendix~\ref{appendix:model_p_under_hyperp_config}.

\noindent \textbf{Inference Latency } We also test the model's inference latency and find no significant increase compared to the baseline. Detailed results and analysis are provided in Appendix~\ref{appendix:inference_latency_analysis}.

\section{Related Works}
\subsection{LM-based TTS models}
Significant progress has been made in using large language models (LLMs) for TTS tasks. Early breakthroughs include VALL-E\cite{wang2023neural}, which pioneered the use of autoregressive and non-autoregressive LMs to predict discrete speech tokens from text or phonemes. Extensions like VALL-E X \cite{zhang2023speak} enable cross-lingual synthesis, and Spear-TTS \cite{kharitonov2023speak} supports multi-speaker TTS with minimal supervision. Recent TTS systems combine AR models with components like diffusion \cite{borsos2023audiolm,lajszczak2024base,anastassiou2024seed} to improve speech quality and control. In contrast, single-stage systems like MELL-E \cite{meng2024autoregressive} and KALL-E \cite{zhu2024autoregressive} avoid this issue but rely on continuous acoustic features, which can hinder large-scale training.



\subsection{GFlowNets}
Generative Flow Network (GFlowNet) \cite{bengio2021flow, bengio2023gflownet} is a probabilistic framework that learns amortized policies to diversely sample structured objects (e.g., molecules, graphs) with probabilities proportional to a predefined reward function, bridging reinforcement learning, generative modeling, and probabilistic inference \cite{zhang2022generative, zhang2022unifying, pan2023better, tiapkin2024generative}. Numerous studies have extended GFlowNets, including connections to variational inference \cite{malkin2022gflownets,zimmermann2022variational}, the integration of intermediate rewards \cite{pan2022generative}, and applications with diffusion models \cite{garipov2023compositional}. Currently, GFlowNets have been applied across a wide range of fields including but not limited to scientific discovery (e.g., molecular design) \cite{jain2023gflownets, koziarski2024rgfn, ghari2023generative}, combinatorial optimization \cite{zhang2023robust, zhang2023let, kim2024ant, hu2024amortizing}, diffusion alignment\cite{venkatraman2024amortizing, zhang2024improving, liu2024efficient}, domain adaptation \cite{zhu2023generalized}, and phylogenetic inference \cite{zhou2023phylogfn}.

\section{Conclusion}
In this work, we propose GOAT, a novel post-training framework leveraging GFlowNets to mitigate hallucinations in LM-based TTS models. By reformulating autoregressive speech generation as a trajectory flow optimization task, we tailor enhanced SubTB and internal reward to align the model’s output distribution toward high-confidence sequences through reward-proportional probability flows. Extensive experiments on multilingual benchmarks demonstrate that GOAT can achieve efficient hallucination suppression without any reliance on high-quality datasets or huge computational resources.
We believe this work offers an inspiring solution for hallucination mitigation in autoregressive speech generation.


\section{Limitations}
We initially analyzed hallucinations in LM-based TTS and proposed GOAT to mitigate them. However, hallucination phenomena are more nuanced than uncertainty alone can capture, so GOAT currently addresses only a subset of cases. To date, our evaluation has focused on CosyVoice 2, a representative LM-based TTS paradigm, laying the groundwork for broader validation. Future work will both extend GOAT to cover a broader range of hallucinations and assess its effectiveness across diverse LM-based architectures.

\section{Acknowledgments}
This research is supported by the National Natural Science Foundation of China under Grant No. 62376071.


\bibliography{cite}
\bibliographystyle{acl_natbib}
\newpage
\appendix

\section{Details of Hallucination Analysis in LM-based TTS models}

\subsection{Modality-Specific Hallucination Detection in Speech}
\label{sec:Modality-Specific Hallucination Detection in Speech}

The investigation of hallucination detection methods remains a critical research topic. While numerous hallucination detection methods have been proposed for text generation tasks, most of them are tailored specifically to the textual modality. However, significant gaps exist between text and speech modalities in token sequence generation. Speech inherently exhibits lower information density and longer token sequences compared to text. 

Crucially, individual text tokens often carry semantic meaning, whereas speech tokens typically lack intrinsic interpretability and must be aggregated to convey coherent information. For instance, a ten-word utterance may require hundreds of acoustic tokens for representation. 

Furthermore, the mapping relationship between token sequences and outputs is fundamentally different across modalities. 
Text generation allows for high-dimensional semantic space mappings, where a single proposition can be expressed through syntactic reordering, lexical substitution, or pragmatic adjustments. In contrast, 
speech synthesis tasks, such as zero-shot generation, impose dual constraints on acoustic token sequences: (1) strict alignment with the symbolic representation of target text (e.g., phonetic/phonological features), and (2) consistency with prosodic representations (e.g., timbre, rhythm, emotion) derived from reference audio. This results in the necessity for almost all tokens in the speech token sequence to be subject to stringent constraints and treated uniformly, requiring comprehensive analysis and optimization across all tokens.
This narrows the feasible solution space for speech generation, requiring models to optimize within tighter acoustic-linguistic boundaries.

In light of the unique characteristics of token sequences in the speech modality, it is necessary to develop specialized methods for analyzing hallucinations in speech. Recent researches pointed out the promising relationship between uncertainty and hallucination in LLM-driven text generation systems\cite{huang2024survey, ma2025estimating}. 

Specifically, in the token generation process of LLMs, output tokens are typically sampled from probability distributions at each decoding step. These distributions inherently reflect the model’s confidence in its predictions: a concentrated distribution (i.e., dominant probabilities assigned to a small subset of tokens) indicates high certainty in the generated token. Conversely, a uniform or dispersed distribution signals ambiguity in decision-making, often correlating with increased uncertainty, which indicates a higher risk of hallucinations that may propagate errors in subsequent steps\cite{zhang2023enhancing, yoffe2024debunc}. 

\subsection{Hallucination Detection Methods}
\label{appendix:hdm}

\begin{figure}[t!]
    \centering
    \includegraphics[width=0.9\linewidth]{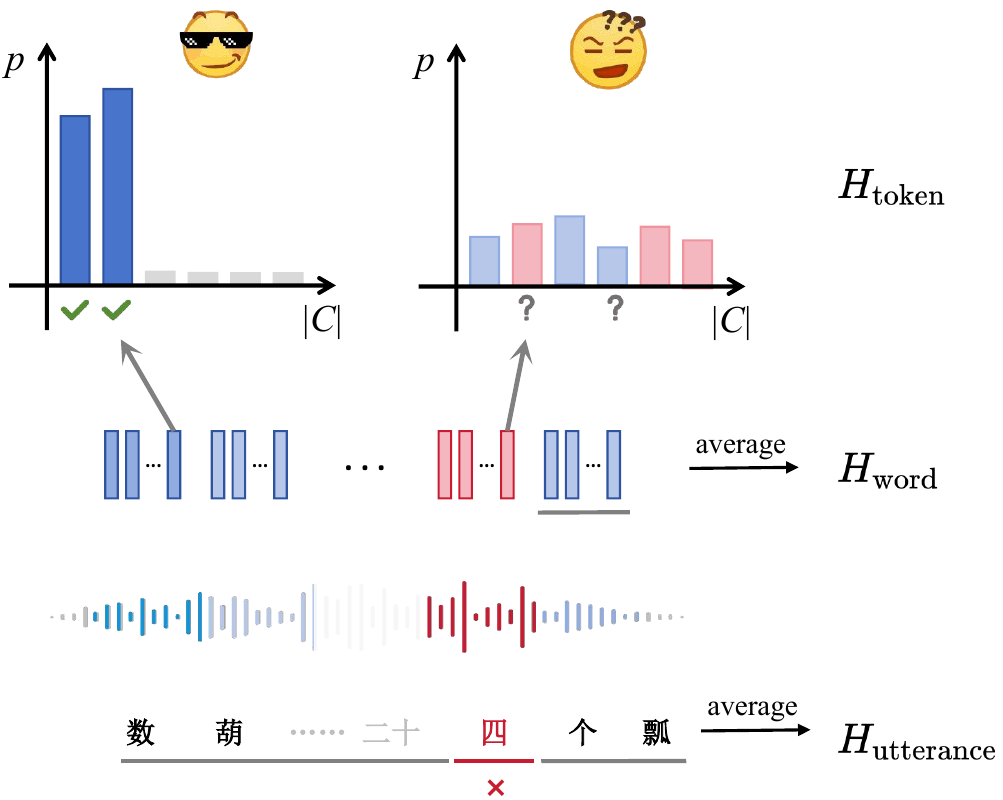}
    \caption{Model uncertainty quantification methods on token, character and utterance levels by entropy.}
    \label{fig:uncertainty_analysis}
\end{figure}

Given huge modality gaps before, we adopt entropy—a more generalizable metric—to detect hallucinations in LM-based TTS models. Let $\mathcal{M}$ denote the pre-trained LLM component in an LLM-based TTS model, with a tokenizer vocabulary $C = \{\tau^1, \tau^2, \dots, \tau^{|C|}\}$, where $|C|$ represents the vocabulary size. The model generates a probability distribution $P_{a_t} = \{p_{\tau^1}, p_{\tau^2}, \dots, p_{\tau^{|C|}}\}$ for the next token at timestep $t$, conditioned on the input prompt token sequence $\mathbf{q}$ and the previously generated sequence $\mathbf{a}_{t-1} = a_1a_2 \cdots a_{t-1}$. Here, $p_{\tau^i}$ (for $i \leq |C|$) denotes the probability of token $\tau^i$, and this process continues until the model samples a termination token. The entropy, i.e., the uncertainty $H_{\text{token}}(P_{a_t})$ of the probability distribution at timestep $t$ is defined as follows: 
\begin{align}
   \label{eq:entropy2}
    H_{\text{token}}(P_{a_t}) = -\sum_{i=1}^{|C|} p_{\tau^i} \log p_{\tau^i} 
\end{align}

Equation (\ref{eq:entropy2}) represents the uncertainty analysis at the token level. However, in speech token sequences, multiple tokens are often required to represent the pronunciation of a single character. For a character $W_{ij}$ composed of the token sequence $a_i, a_{i+1}, \dots, a_j$, the uncertainty $H_{\text{character}}(W_{ij})$ is computed as the average uncertainty of the tokens constituting the character. Similarly, for an utterance $S$ with a token sequence length of $|S|$, the uncertainty $H_{\text{utterance}}(S)$ is calculated as the mean uncertainty of all tokens in the utterance. The formulas for character-level and utterance-level uncertainty are as follows:
\begin{align}
    H_{\text{character}}(W_{ij}) &= \frac{1}{j-i} \sum_{k=i}^j H_{\text{token}}(P_{a_k}) \\
    H_{\text{utterance}}(S) &= \frac{1}{|S|} \sum_{k=1}^{|S|} H_{\text{token}}(P_{a_k}) \label{eq:utterance_uncertainty2}
\end{align}

An illustration of our detection method is provided in Figure~\ref{fig:uncertainty_analysis}. In the context of LLM generation tasks, entropy can be interpreted as a measure of the model's uncertainty regarding its output. When the model is highly confident in predicting a specific token, the resulting probability distribution exhibits low entropy, indicating a stable generation process. Conversely, a high-entropy distribution suggests that the model is uncertain among multiple candidate tokens, leading to increased sampling variability and potential susceptibility to randomness-induced fluctuations.

\subsection{Detailed Fine-grained Hallucination Phenomena Analysis with Uncertainty}
\label{appendix:fine-grained_analysis}

Based on our experimental results at the utterance level, we observe that using CER and entropy as evaluation proxies—along with external factors such as ASR errors—inevitably introduces noise. This helps explain why the obtained correlation coefficients (0.636 and 0.649) are only moderately strong. To enable a more fine-grained analysis, we further examine token-level uncertainty during the computation of utterance-level uncertainty. We also visualize token-wise uncertainty using line plots for each test utterance, allowing for detailed inspection of local variations.
By leveraging temporal alignment from ASR outputs, we identified token subsequences corresponding to individual characters. For character-level uncertainty, we computed the average uncertainty of constituent tokens and generated character-wise uncertainty line plots. 

To address ASR limitations (e.g., insensitivity to prosodic/pause errors), we manually validate a subset of hallucination cases through human evaluation of speech-text alignment. We categorized hallucinations into five primary types:

\begin{figure*}[t!]
    \centering
    \includegraphics[width=0.9\textwidth]{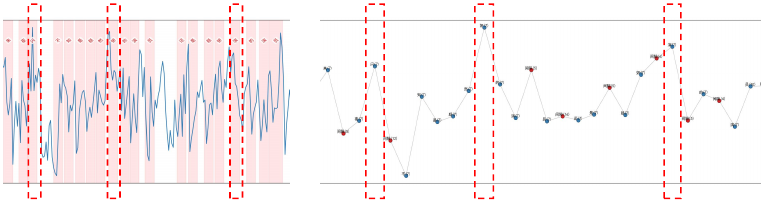}
    \caption{One classic case of Mispronounced/Incorrect characters, with red dashed boxes highlighting error regions.}
    \label{fig:mispronounced_incorrect_word}
\end{figure*}

\textbf{Mispronounced/Incorrect characters } In these cases, the model generates speech containing lexically incorrect terms or phonetic errors. Since each character in speech token sequences is represented by multiple tokens, such errors manifest as faulty subsequence generation, indicating hallucinatory behavior. As is visualized in Figure~\ref{fig:mispronounced_incorrect_word}, both token-level and character-level uncertainties are significantly elevated in these cases.

\begin{figure*}[t!]
    \centering
    \includegraphics[width=0.9\textwidth]{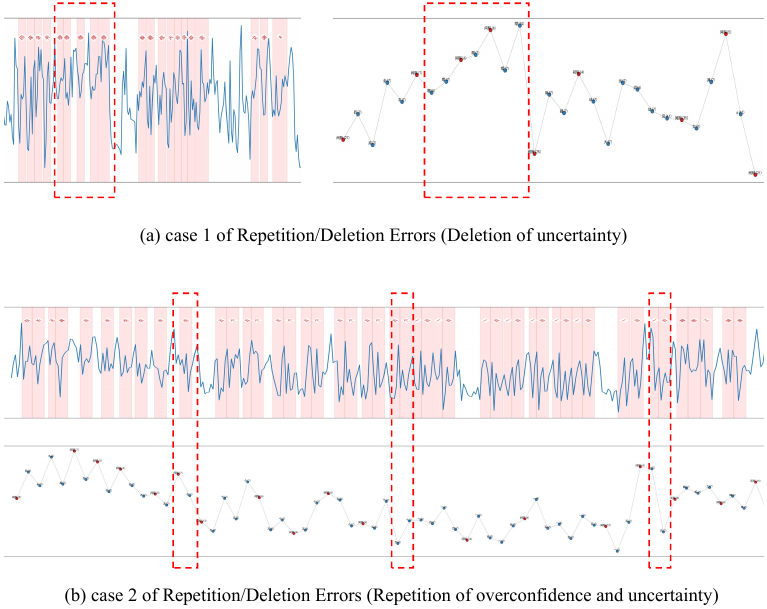}
    \caption{Two classic cases of Repetition/Deletion Errors, with red dashed boxes highlighting error regions. (a) Deletion error case lacking one time of the phonetic token 'ying'. (b) The first and third hallucination examples exhibit high uncertainty in multiple repetition errors, while the second repetition case demonstrates overconfidence.}
    \label{fig:repetition_deletion_error}
\end{figure*}

\textbf{Repetition/Deletion Errors } These kind of errors occur when the model fails to handle repetitive text, resulting in missing or extra character repetitions. As is visualized in Figure~\ref{fig:repetition_deletion_error}, some errors arise from high uncertainty (e.g., skipped characters in complex repetitions), while others stem from overconfidence—where uncertainty decreases despite errors.

\begin{figure*}[t!]
    \centering
    \includegraphics[width=0.9\textwidth]{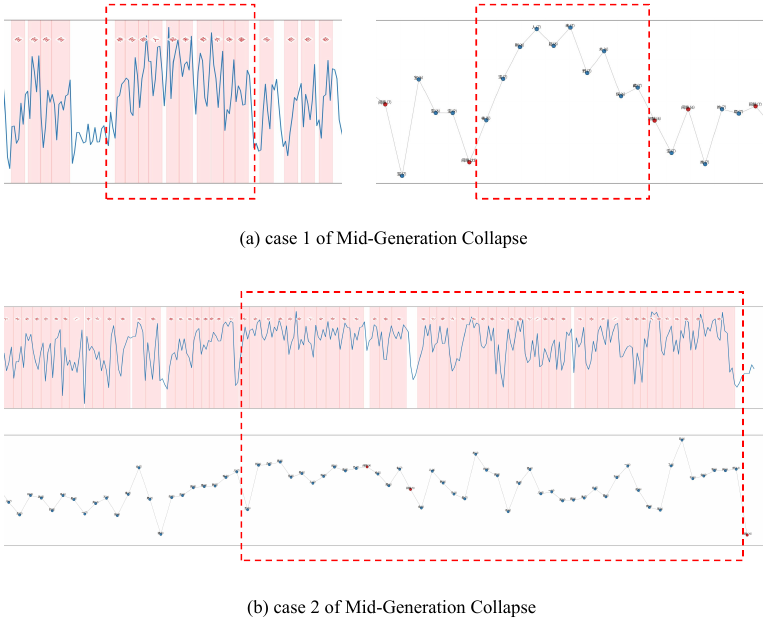}
    \caption{Two classic cases of Mid-Generation Collapse, with red dashed boxes highlighting error regions. (a) A mid-generation collapse occurred in the speech token sequence (b) Starting from a certain step in the middle, the generated content becomes entirely incoherent until termination.}
    \label{fig:mid-generation_collapse}
\end{figure*}

\textbf{Mid-Generation Collapse } This phenomenon represents severe hallucinations where speech becomes incoherent after a certain point. Generated content is often unrecognizable and lacks logical structure, reflecting profound failures in autoregressive generation. As is visualized in Figure~\ref{fig:mid-generation_collapse}, consistently high uncertainty is observed during these collapse phases.

\begin{figure*}[t!]
    \centering
    \includegraphics[width=0.9\textwidth]{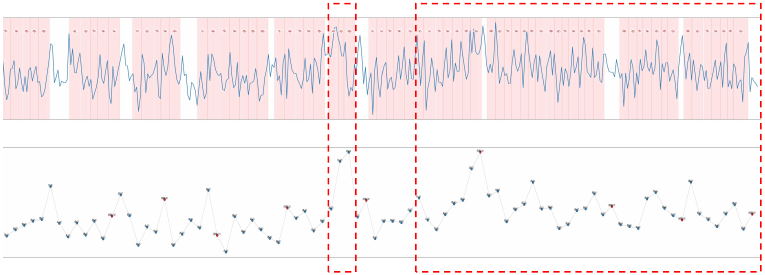}
    \caption{One classic case of Content Divergence, with red dashed boxes highlighting error regions. The first error region includes character mispronunciations, and the second large region is inconsistent with target text although preserving semantically related contents.}
    \label{fig:semantic_divergence}
\end{figure*}

\textbf{Content Divergence } This involves semantically meaningful but textually mismatched content, typically in long sentences.As is visualized in Figure~\ref{fig:semantic_divergence}, while the generated speech remains intelligible and semantically related to targets, textual alignment breaks down, which represents the occurrence of hallucinations. Relatively large uncertainty is observed despite the superficial coherence of outputs.

\begin{figure*}[t!]
    \centering
    \includegraphics[width=0.9\textwidth]{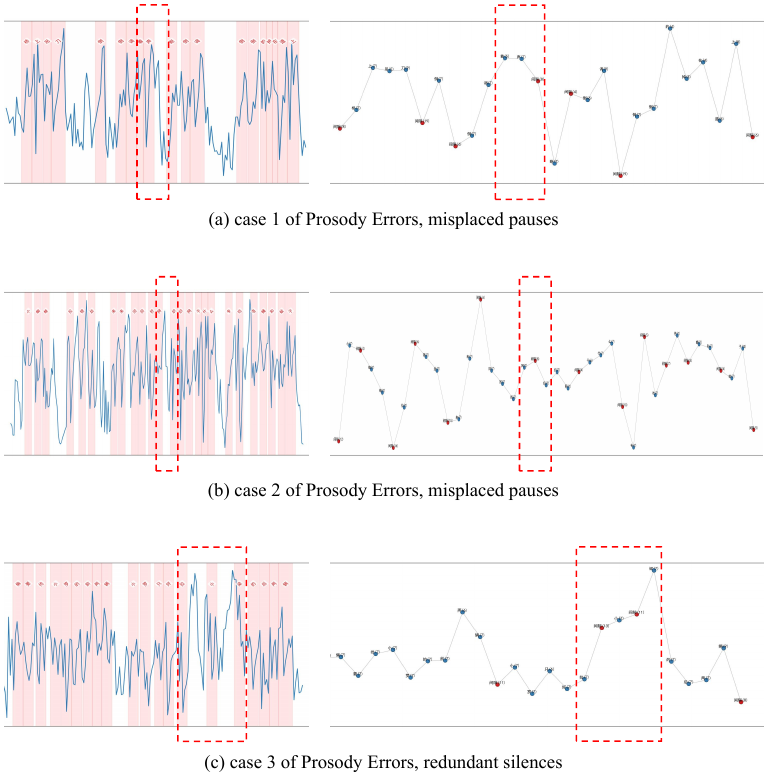}
    \caption{Three classic cases of Prosody errors, with red dashed boxes highlighting error regions. (a) The error region contains two characters with a gap that should be positioned between them. (b) In this error region, the separated characters should be coherently pronounced, followed by a brief pause. (c) Both larger gaps are redundant silences and should not occur in this error region.}
    \label{fig:prosody_errors}
\end{figure*}

\textbf{Prosody Errors } Prosody errors include misplaced pauses or redundant silences that may impair content comprehension. As is visualized in Figure~\ref{fig:prosody_errors}, unlike other categories, no consistent uncertainty pattern emerges for prosodic hallucinations. Only part of high uncertainty tokens are observed in some pauses or characters nearby. This complexity suggests the need for specialized metrics tailored to prosodic representation learning.

\section{Preliminaries of GFlowNets}
\label{sec:appendix_gflownets}

Since the autoregressive sampling process under consideration is unidirectional (i.e., forward-only), we focus on the forward process of GFlowNets in this section. Following the mathematical notation in \cite{malkin2022trajectory}, we define a directed acyclic graph (DAG) $G = (\mathcal{S}, \mathcal{A})$, where $\mathcal{S}$ represents the set of \textit{states} (vertices), and $\mathcal{A}$ denotes the set of \textit{actions} (edges), such as $(u \to v) \in \mathcal{A}$ for $u, v \in \mathcal{S}$. The unique \textit{initial state} $s_0 \in \mathcal{S}$ has no incoming edges. All \textit{terminal states} (with no outgoing edges) form the set $\mathcal{X}$. A \textit{trajectory} $\tau$ is a sequence of states being defined as $\tau = (s_m \rightarrow s_{m+1} \rightarrow \cdots \rightarrow s_n)$, where $\forall i = m, m+1, \dots, n-1, (s_i \to s_{i+1}) \in \mathcal{A}$. If a trajectory satisfies $s_m = s_0$ and $s_n \in \mathcal{X}$, it is termed \textit{complete}. The set of all complete trajectories is denoted $\mathcal{T}$.

To construct trajectories, GFlowNets sample decisions from a \textit{forward policy} $P_F(\cdot|s)$, where $s \in \mathcal{S}$ denotes the current state. This policy iteratively transitions from one state to the next until reaching a terminal state $s_n$. Consequently, the forward policy implicitly defines a distribution over all successor states for non-terminal states $s \in \mathcal{S}$. For a complete trajectory $\tau \in \mathcal{T}$, the probability distribution is given by:  
\begin{align}
    P_F\big(\tau = (s_0 \to \cdots \to s_n)\big) = \prod_{i=0}^{n-1} P_F(s_{i+1} \mid s_i),
\end{align}
where all trajectories sampled from the forward policy satisfy the Markov property: the distribution of any non-initial state depends solely on its immediate predecessor. Thus, the objective of GFlowNets is to sample a complete trajectory from the forward policy, which corresponds to a terminal state $x \in \mathcal{X}$. The marginal likelihood of $x$ is defined as the sum of probabilities of all trajectories terminating at $x$:  $\sum_{\substack{\tau = (s_0 \to \cdots \to s_n = x)}} P_F(\tau)$

Given a non-trivial non-negative target reward function $R: \mathcal{X} \to \mathbb{R}_{\ge 0}$, the objective of GFlowNets is to learn a forward policy $P_F$ such that the probability of sampling any terminal state $x$ is proportional to its reward $R(x)$. In other words, the marginal likelihood distribution of terminal states should align with the reward function (which need not be normalized). Formally, there exists a non-negative constant $Z$ satisfying:  
\begin{align}
\label{eq:reward_proportionality}
    R(x) = Z \sum_{\substack{\tau = (s_0 \to \cdots \to s_n = x)}} P_F(\tau) \quad \forall x \in \mathcal{X}.
\end{align}
Analogous to its name, the GFlowNet sampling process can be visualized through a water-flow analogy: Define a \textit{trajectory flow} $F: \mathcal{T} \to \mathbb{R}_{\ge 0}$, where probability "flows" along trajectories like water. The forward policy $P_F$ is then derived from $F$ as:  
\begin{align}
    P_F(\tau) = \frac{1}{Z_F} F(\tau), \quad Z_F = F(s_0) = \sum_{\tau \in \mathcal{T}} F(\tau)
\end{align}
where $Z_F$ normalizes the total flow. Notably, the absolute scale of $F$ is arbitrary—it can be rescaled without affecting the induced policy. For any state $s \in \mathcal{S}$, the \textit{state flow} $F(s)$ is defined as the total flow passing through $s$: $F(s) = \sum_{ s \in \tau} F(\tau)$. Under this framework, the goal of GFlowNets is to approximate a flow $F$ that satisfies:  
\begin{align}
    \label{eq:flow_normalization2}
    F(x) = R(x) \quad \forall x \in \mathcal{X}.
\end{align}
Equations ($\ref{eq:reward_proportionality}$) and ($\ref{eq:flow_normalization2}$) are equivalent in objective. If this condition holds, the total flow $Z_F$ equals the normalization constant $Z = \sum_{x \in \mathcal{X}} R(x)$, representing the aggregate reward across all terminal states.

\section{Datasets}
\label{appendix:datasets}


\textbf{Training Dataset } To evaluate the generalizability of our approach, we conducted training on two major languages, Chinese and English. For English, we used the train-clean-100 subset of the LibriTTS\cite{zen2019libritts} corpus as the source of prompts and target text during training, which contains speech data from 247 speakers with good speakers generalizability. Specifically, we randomly selected 1,000 text-to-speech pairs from the train-clean-100 subset to serve as prompts for training, and then randomly selected another 1,000 text samples as target text for synthesis. These target texts were combined with the previously selected prompts to form the training dataset. For Chinese, we used the premium subset of WenetSpeech4TTS\cite{ma2024wenetspeech4tts}, which also contains a large number of speakers, as the source of prompts and target text, following the same procedure as for the English dataset. We also trained our model on a mixed dataset. Specifically, keeping the total training data volume unchanged (still 1,000 samples), we randomly sampled 500 instances from each of the Chinese and English datasets, which together formed a combined training dataset for model training.


\textbf{Evaluation Dataset} Following the methodology of CosyVoice 2, we used the official seed-tts-eval test set\cite{anastassiou2024seed}, which contains three sub datasets, including approximately 1,000 English test samples (\textit{test-en}) from the Common Voice dataset, and 2,000 Chinese test samples (\textit{test-zh}) from the DiDiSpeech-2 dataset. Additionally, there are around 400 challenging samples in the case \textit{test-hard} subset, which includes difficult synthesis targets such as text repetition, tongue twisters, phonetically similar texts, and commonly mispronounced characters. We use the same testing methodology and dataset as CosyVoice 2 to demonstrate the improvements made by our approach in enhancing model performance.


\section{Implementation Details}
\label{appendix:implementation_details}

In this experiment, we provide two sets of hyperparameter configurations, which differ primarily in the total number of steps in the Learning Rate Optimization (LRO) process. Specifically, for all experiments, the total training rounds are fixed at 15, corresponding to approximately 3500 global steps. The entire training process takes about 7 hours on four NVIDIA H100 Hopper GPUs, with each GPU utilizing around 70 GB of memory. As described in Section~\ref{sec:reward_hacking_suppression}, to balance model performance and training stability while suppressing reward hacking, we set the minimum reward temperature to 0.825 and the maximum learning rate to 1e-5. Two configurations with slight differences in LRO are provided, where the warm-up phase is fixed at 20 steps. We adjust the remaining cosine annealing steps to achieve total LRO steps of 1500 and 2500, corresponding to more stable training and stronger model performance respectively.

It is worth noting that, given the properties of GOAT, we speculate that it should still perform effectively even with weaker or even no prompts. Reducing the prompt length or the target synthesis text length could significantly reduce GPU memory usage. We leave the exploration of more efficient training strategies for GOAT to future work.

During the inference stage, we employ the Repetition Aware Sampling (RAS) method using the default configuration from CosyVoice 2. Specifically, the top-p value is set to 0.8, top-k is set to 25, the repetition penalty window size (win\_size) is 10, and the repetition penalty coefficient (tau\_r) is 0.1.

\section{Metrics Details}
\label{appendix:metrics}
For the Chinese dataset, we deploy the Paraformer-zh ASR model\cite{gao2022paraformer} to detect the content of the synthesized speech and compute the character error rate (CER) by comparing it with the target text. For the English dataset, we use the Whisper-large V3 model\cite{radford2023robust} for speech recognition and calculate the corresponding word error rate (WER). For speaker similarity (SS), we uniformly use the CAM++ model\cite{wang2023cam++} to extract speaker feature vectors from both the prompt audio and the generated audio, then compute the cosine similarity of the speaker embeddings, and finally average the results to represent the speaker similarity evaluation. Lastly, for speech quality, we use the objective evaluation metric UTMOS\cite{saeki2022utmos} to measure the synthesized speech, which is one of the commonly used evaluation metrics for assessing the naturalness and quality of synthesized speech.

\section{Model Performance under Different Configuration}
\label{appendix:model_p_under_hyperp_config}

We evaluate models with different parameter settings. Specifically, we use the CV2-GOAT-en(1500S) training setup, varying one hyperparameter at a time, and test on the \textit{test-en} dataset. For stably trained models, we measure WER using the last three epochs; for models affected by reward hacking, we evaluate the first three epochs before the impact by reward hacking. The performance evaluation and analysis for different settings are presented below.

\begin{table}[t!]
    \small
    \centering
    \begin{tabular}{l c c c}
        \toprule
        & Epoch -3 & Epoch -2 & Epoch -1 \\
        \midrule
        Ours & 2.270 & 2.174 & 2.160 \\
        w/o RTD & 2.753 & 2.705 & 2.750 \\
        w/o LRO & 5.453 & 3.485 & 2.643 \\
        \bottomrule
    \end{tabular}
    \caption{\label{tab:ablation_configuration_comparison}Performance comparison on ablation study}
\end{table}



\textbf{w/o RTD } As shown in Table~\ref{tab:ablation_configuration_comparison}, although discarding RTD yields a seemingly more stable training process, the performance of the resulting model is not particularly good, which demonstrates that the RTD strategy is crucial for promoting training convergence and distribution alignment.

\textbf{w/o LRO } As shown in Table~\ref{tab:ablation_configuration_comparison}, for models trained without LRO, reward hacking occurred before convergence, and their performance over the final three epochs fell short of expectations. This observation demonstrate the efficacy of LRO in guiding model training. 

\begin{table}[t!]
    \small
    \centering
    \begin{tabular}{l c c c}
        \toprule
        Min Reward Temp & Epoch -3 & Epoch -2 & Epoch -1 \\
        \midrule
        0.925 & 3.240 & 3.224 & 3.053 \\
        0.875 & 2.569 & 2.681 & 2.831 \\
        0.825 (Ours) & 2.270 & 2.174 & 2.160 \\
        0.775 & 2.163 & 2.285 & 2.855 \\
        0.725 & 2.589 & 2.138 & 2.697 \\
        \bottomrule
    \end{tabular}
    \caption{\label{tab:min_reward_temp}Performance Comparison with different minimum reward temperature (Min Reward Temp)}
\end{table}

\textbf{Minimum Reward Temperature } Table~\ref{tab:min_reward_temp} shows that a higher minimum reward temperature ensures very stable training but yields suboptimal performance, whereas a lower minimum temperature destabilizes training, leading to premature reward hacking and large performance fluctuations. Our configuration balances performance and stability.

\begin{table}[t!]
    \small
    \centering
    \begin{tabular}{l c c c}
        \toprule
        LRO Steps & Epoch -3 & Epoch -2 & Epoch -1 \\
        \midrule
        1500 (Ours) & 2.270 & 2.174 & 2.160 \\
        2500 (Ours) & 2.557 & 2.336 & 2.133 \\
        3500 & 2.693 & 2.628 & 2.182 \\
        \bottomrule
    \end{tabular}
    \caption{\label{tab:lro_steps}Performance comparison with different LRO steps}
\end{table}

\textbf{LRO steps } Table~\ref{tab:lro_steps} shows that with 3500 LRO steps, the model undergoes premature reward hacking, fails to align the distribution, and thus suffers degraded performance. In contrast, 1500 and 2500 steps respectively yield stable convergence and a controlled trade-off of stability for improved performance.

\begin{table}[t!]
    \small
    \centering
    \begin{tabular}{l c c c}
        \toprule
        Max Learning Rate & Epoch -3 & Epoch -2 & Epoch -1 \\
        \midrule
        $5\mathrm{E}-5$ & 3.901 & 3.067 & 2.776 \\
        $1\mathrm{E}-5$ (Ours) & 2.270 & 2.174 & 2.160 \\
        $5\mathrm{E}-6$ & 2.791 & 2.788 & 3.082 \\
        \bottomrule
    \end{tabular}
    \caption{\label{tab:max_lr}Performance comparison with different maximum learning rate}
\end{table}

\textbf{Maximum Learning Rate } Regarding the maximum learning rate, as shown in Table~\ref{tab:max_lr}, setting it too high provokes premature reward hacking, curtailing convergence and degrading performance. In converse, setting it too low leaves optimization underutilized. Our chosen rate strikes a balance, mitigating reward hacking while maintaining strong model performance.

\section{Inference Latency Analysis}
\label{appendix:inference_latency_analysis}

\begin{figure}[t!]
    \centering
    \includegraphics[width=0.9\linewidth]{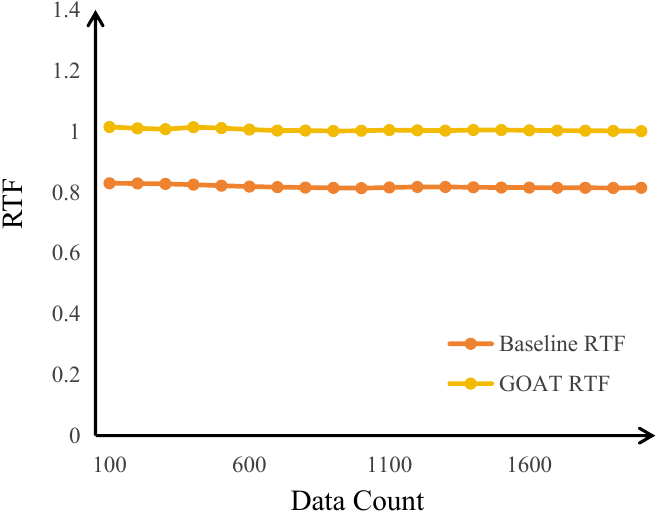}
    \caption{RTF compared with baseline}
    \label{fig:rtf}
\end{figure}

We use Real-Time Factor (RTF) for inference latency analysis. RTF is a widely used metric in the speech generation field to measure the efficiency of a model in terms of its processing speed. It is defined as the ratio between the time taken for the model to generate speech and the duration of the generated audio. Specifically, the RTF is calculated as:
\begin{equation}
    \text{RTF} = \frac{\text{Generation Time}}{\text{Audio Duration}}
\end{equation}
An RTF of 1.0 indicates that the model generates speech in real-time, while an RTF greater than 1.0 implies that the model takes longer than real-time to generate the speech. A lower RTF is desirable, as it reflects faster generation, which is crucial for practical deployment of speech synthesis systems.

We evaluated the RTF on \textit{test-zh} dataset of 2000 samples, accumulating the results in batches of 100. As shown in Figure~\ref{fig:rtf}, the GOAT-trained model does not incur significant additional inference latency compared to the baseline, with only the delay introduced by LoRA. It still maintains real-time generation capability on a V100 GPU.

\end{document}